\documentclass[aps,prb,preprint,showpacs]{revtex4}

\usepackage{amsmath}
\usepackage{graphicx}
\usepackage{setspace}
\usepackage{bm}
\usepackage{dcolumn}
\usepackage{ulem}

\begin{document}

\title{Mathematical model of flux relaxation phenomenon}

\author{Rongchao Ma} \email{marongchao@yahoo.com}
\affiliation{Department of Physics, University of Alberta, Edmonton, Alberta, Canada}

\date{\today}

\begin{abstract}
The investigations on the flux relaxation phenomenon of a type-II superconductor are important because they provide the information about the flux pinning ability and current-carrying ability of the superconductor. However, a unified theory of flux relaxation is currently unavailable. Here I present a general mathematical model of the flux relaxation. In this model, I proposed a series expansion to the activation energy and derived a general formula for the current decay behavior. In the light of these formulas, I can analyze the experimental data on the current decay behavior and then calculate the activation energy of a vortex system without subjecting to any special conditions. The results are accurate for the current decay measurements from a $Bi_2Sr_2CaCu_2O_{8+x}$ superconductor.
\end{abstract}


\maketitle

\section{Introduction}

The persistent current (or trapped magnetic field) in a type-II superconductor decays because the vortex lines jump between adjacent pinning centers spontaneously due to thermal activation \cite{Landau}, quantum tunneling \cite{Koren,Nicodemi,Hoekstra} or mechanical vibration. Flux relaxation has dependency on activation energy and temperature. The activation energy is a function of current because the Lorentz force of a current reduces the activation energy. By proposing a detailed current dependent activation energy, one can obtain the corresponding current decay model. In low temperature superconductors, the strong pinning dominates and current decay usually shows logarithmic behavior.\cite{Tinkham} This can be explained by Anderson's flux creep theory in which he assumed an activation energy with linear current dependence.\cite{Anderson} However, in high temperature superconductors, the weak point pinning dominates and current decay usually shows non-logarithmic behavior.\cite{Tinkham} To address this phenomenon, a number of theoretical models were proposed, such as collective creep theory \cite{Feigel'man1} and vortex glass theory \cite{Fisher}, both of which give an activation energy with nonlinear current dependence \cite{Feigel'man2,Malozemoff}. Other theoretical models were also proposed for the non-logarithmic behavior of current decay, like bulk-surface pinning transition \cite{Burlachkov,Chikumoto,Weir} and flux redistribution \cite{Gurevich1}.

The work described here is motivated by the fact that the flux relaxation phenomenon is currently described with multiple models. The experimental confirmation of these theories are limited to the special cases.\cite{Kim1,Beasley,Miu,Yang,Reissner} In practice, the temperature and current are arbitrary. The discrepancy between the theoretical predications and experimental results are large. Thus, it is highly desirable to search for a general theoretical model which gives an arbitrary description of the activation energy and corresponding current decay behavior without subjecting it to any restraint conditions. The studies on flux relaxation are important because the flux relaxation influences the current carrying ability and stability of a type-II superconductor; consequently, it limits the possible applications of the superconductor. The information about the flux relaxation is also important in understanding the pinning mechanism and vortex structure in a type-II superconductor.

In this paper, I investigated the general expressions of flux relaxation phenomenon from a mathematical framework. First, I expanded the activation energy as a Taylor series of current density and obtained the corresponding flux relaxation model using the inverse series methods. Next, I discussed the possible physical meaning of various parameters. Finally, I studied the applied conditions of the quadratic activation energy and linear activation energy.

\section{Model}

Flux relaxation can be described by hopping rate, $R \propto e^{-U_a/kT}$, where $U_a$ is the activation energy of vortices, $k$ is Boltzmann constant and $T$ is temperature.\cite{Tinkham} The hopping rate of the vortices is increased by Lorentz force $f_L = j \times B $, where $j$ is current density and $B$ is magnetic field. Therefore, $U_a$ is a decreasing function of $j$.(Ref.~\onlinecite{Tinkham}) The thermal hopping of vortices causes a fall in $j$, and $j$ is then a function of time. By proposing a detailed $j$ dependent $U_a(j)$, one can obtain the time evolution of $j(t)$. This suggests that an important step in studying the flux relaxation phenomenon is to search for a detailed $j$ dependent $U_a(j)$. Based on different physical considerations \cite{Anderson,Fisher,Feigel'man2,Zeldov,Malozemoff}, physicists obtained a number of $U_a(j)$. Here I used an approach mainly based on mathematical consideration.

First, I argue that the activation energy $U_a(j)$ must be a nonlinear function of current density $j$ because of the elasticity of vortices and interaction between vortices.\cite{Beasley} A vortex deforms under a driving force. It may yield under a strong driving force and show plastic behavior \cite{Kierfeld,Abulafia}. In this case, the strain energy of a vortex is a nonlinear function of $j$. On the other hand, the interaction energy between vortices is $U_{int} = \frac{\Phi_0^2}{8\pi^2\lambda^2}K_0\left(\frac{r}{\lambda}\right)$, where $r$ is the distance between two vortices, $\lambda$ is the penetration depth and $K_0$ is a zeroth order Hankel function with an imaginary argument.\cite{Tinkham} Function $K_0$ causes a nonlinear response in $U_a(j)$ with respect to $j$. Therefore, the $U_a(j)$ of a real vortex system must be a nonlinear function of $j$.

The exact form of the activation energy is usually unknown because of the complexity of the interaction between vortices. From mathematics we know that series expansion is a good approach for expressing an implicit function. Let us now consider the possibility of expanding the activation energy $U_a(j)$ as a Taylor series with respect to $j$.

For a vortex system, we only need to consider $U_a(j)$ on the closed interval $\left[0, j_c\right]$. The hopping rate, $R \propto e^{-U_a/kT}$, has limited values because a vortex system decays at any circumstances due to the thermal motion of the vortices. It indicates that $U_a(j)$ also has limited values; otherwise, a vortex system does not decay. At vanishing driving force, $U_a(j)$ is the pinning potential $U_c$, i.e., $U_a(j\rightarrow 0)\rightarrow U_c$. $U_c$ is a limited number.\cite{Kes} On the other hand, at critical level, $U_a(j)$ approaches to zero, i.e., $U_a(j\rightarrow j_c) \rightarrow 0$. Thus, $U_a(j)$ is a bounded function for all $j \in \left[0,j_c\right]$. Therefore, in most cases one can expand $U_a(j)$ as a Taylor series of $j$ (at least up to order one, i.e., the linear activation energy). Let us now write out the general expression explicitly:
\begin{equation}
\label{UaGeneral}
U_a(j) = U_c - \sum\limits_{i=1}^n a_i j^i
\end{equation}
where $U_c=U_a(0)$, $ a_1=-U'_a(0)$, $a_2=-U''_a(0)/2!$, $\cdots$, $a_n=-U^{(n)}_a(0)/n!$. Normally, the activation energy $U_a$ has dependence on temperature $T$ and current density $j$. Furthermore, $U_a$ includes the contribution from the elastic deformation and interaction of vortices, which has a dependence on the penetration depth $\lambda$ and coherence length $\xi$. Therefore, the coefficients in Eq.(\ref{UaGeneral}) should be in the form of $U_c(T, \lambda, \xi)$ and $a_n(T, \lambda, \xi)$.

To obtain the current decay model $j(t)$ from Eq.(\ref{UaGeneral}), we need the connection between the activation energy $U_a(j)$ and time $t$. Early studies \cite{Geshkenbein} have shown that, by assuming that flux relaxation is caused by thermal activation, the time evolution of $U_a(j(t))$ (with logarithmic accuracy) is,
\begin{equation}
\label{UvsTime}
U_a(j(t)) = k T ln (1+t/t_0)
\end{equation}
where $t_0$ is a short time scale for the Bean model to be formed.\cite{Feigel'man2} This is a macroscopic quantity depending on sample size. Substituting Eq.(\ref{UvsTime}) into Eq.(\ref{UaGeneral}), we have
\begin{equation}
\label{WvsJ}
w(t)= \sum\limits_{i=1}^n a_i j^i(t)
\end{equation}
where
\begin{equation}
\label{FunctionW}
w(t)= U_c - kTln(1+t/t_0).
\end{equation}

Now our task is to find out the inverse function $j(w(t))$ from Eq.(\ref{WvsJ}). First, let us expand $j(w(t))$ as a series of $w(t)$,
\begin{equation}
\label{JvsW}
j(w(t)) = \sum\limits_{i=1}^n b_i w^i(t)
\end{equation}

Employing Cauchy's formula \cite{Morse} from complex analysis, one can easily obtained the coefficients $b_i$'s \\
\begin{equation}
\label{bns}
\begin{aligned}
b_1=&\frac{1}{a_1} \\
b_2=&\frac{1}{a_1^2}\left(-\frac{a_2}{a_1}\right) \\
b_3=&\frac{1}{a_1^3}\left[2 \left(\frac{a_2}{a_1}\right)^2 - \left(\frac{a_3}{a_1}\right)\right] \\
&\cdots\cdots\cdots\cdots \\
b_n=&\frac{1}{a_1^n} \frac{1}{n} \sum\limits_{s,t,u \cdots} (-1)^{s+t+u+\cdots} \left(\frac{a_2}{a_1}\right)^s \left(\frac{a_3}{a_1}\right)^t \cdots \\
    &\cdot \frac{n(n+1)\cdots(n-1+s+t+u+\cdots)}{s!t!u!\cdots}
\end{aligned}
\end{equation}
where $s+2t+3u+\cdots=n-1$. From Eq.(\ref{bns}) we see that only coefficient $a_1$ contributes to the logarithmic decay coefficient $b_1$. The coefficients $a_2, a_3, \cdots, a_n$ result in the non-vanishing coefficients $b_n$ ($n\geq 2$), which contribute to the deviation away from the logarithmic decay.

Since Eq.(\ref{WvsJ}) and Eq.(\ref{JvsW}) have symmetry, if we switch their sequence, the same procedure can be employed to calculate the coefficients $a_n$. Therefore, $a_n$ and $b_n$ have the same symmetry as that of Eq.(\ref{WvsJ}) and Eq.(\ref{JvsW}). By simply doing a commutation to the coefficients $b_n \leftrightarrow a_n $, we have \\
\begin{equation}
\label{ans}
\begin{aligned}
a_n= &\frac{1}{b_1^n} \frac{1}{n}\sum\limits_{s,t,u \cdots} (-1)^{s+t+u+\cdots} \left(\frac{b_2}{b_1}\right)^s \left(\frac{b_3}{b_1}\right)^t \cdots \\
     &\cdot\frac{n(n+1)\cdots(n-1+s+t+u+\cdots)}{s!t!u!\cdots}
\end{aligned}
\end{equation}
where $s+2t+3u+\cdots=n-1$. In practice, we usually first find out the coefficients $b_n$ from experiments, and then calculate the coefficients $a_n$ to obtain $U_a(j)$ from Eq.(\ref{UaGeneral}). 
The $U_a(j)$ obtained in this way includes the contributions from the Lorentz force, deformation of vortices, interaction between vortices and possibly other unknown sources. Thus, $U_a(j)$ is a combined response to $j$. Figure 1 shows that these results are accurate for the experimental data of a $Bi_2Sr_2CaCu_2O_{8+x}$ single crystal.

In Eq.(\ref{UaGeneral}), we expanded $U_a(j)$ as a function of $j$, the terms including $j$ represent the contribution from the driving force. As a result of $j$ being in a superconductor, there will be a corresponding magnetic field $B$ according to the Biot-Savart law. So, we can also expand $U_a$ as a function of the magnetic field $B$ or magnetization $M$ in the same way as we did in Eq.(\ref{UaGeneral}), that is, $U_a(B) = U_c -
\sum\limits_{i=1}^n a_i B^i$, or $U_a(M) = U_c - \sum\limits_{i=1}^n a_i M^i$. Similar to Eq.(\ref{JvsW}), we can obtain the time evolution of the magnetic field $B(t) = \sum\limits_{i=1}^n b_i w^i(t)$, or time evolution of the magnetization $M(t) = \sum\limits_{i=1}^n b_i w^i(t)$.

\section{Discussion}

In Eq.(\ref{UaGeneral}), we expanded the activation energy $U_a(j)$ as a Taylor series of current density $j$. Let us now discuss the possible physical meaning of the coefficients.

1. \textit{Meaning of $U_c$}.---
Eq.(\ref{UaGeneral}) indicates that $U_a(0)=U_c$. The constant $U_c$ is then the activation energy at vanishing driving force, i.e., the pinning potential of a vortex.

The pinning potential $U_c$ is important in determining the irreversibility field $H_i = \frac{E_c}{\mu_0 a_f \nu_0}e^{U_c/k_BT}$, where $E_c$ is the electric field criterion when the current density reduces to zero, $\mu_0$ is the permeability of vacuum, $a_f$ is the flux line spacing and $\nu_0$ is the attempt frequency of the vortex.\cite{Matsushita} On the other hand, $U_c$ can be used to calculate the critical current density $j_c$. From Eq.(\ref{UvsTime}) and Eq.(\ref{FunctionW}) we see that the constraint condition $U_a(j_c)=0$ equals to set $w(t)=U_c$. So, we can obtain the critical current density by replacing the $w(t)$ in Eq.(\ref{JvsW}) with $U_c$, that is, $j_c =  \sum\limits_{i=1}^n b_i U_c^i$.

2. \textit{Meaning of $a_n$} (n$\geq$2).---
From mechanics we know that elastic reaction is represented by the quadratic term of independent variables, while inelastic reactions are represented by the higher order terms (above two) of independent variables. Let us now prove that the activation energy of a elastic vortex system has quadratic current dependence.

Consider an elastic vortex which is subjected to a uniformly distributed Lorentz force $f_L$, the vortex undergoes deformation and adjusts its position to minimize the total energy because of the elasticity. For simplicity, let us consider the vortex as a uniform object. The elastic strain energy due to the bending of the vortex is $U_e(j)= \int^{L}_{0}\frac{M^2}{2EI}dx$, where $M$ is the internal moment, $E$ is the modulus of elasticity, $I$ is the moment of inertia, $L$ is the length of the vortex under consideration and $x$ is the coordinate established along the vortex.\cite{Hibbeler} Under the uniformly distributed load $f_L$, the equilibrium condition is, $M+x\left(\frac{x}{2}\right)f_L=0$. Assuming that the deformation of the vortex $\delta$ is very small compared with the distance between two adjacent pinning centers $D$ ($\delta<<D$), one can find the elastic strain energy due to the bending of the vortex: $U_e(j) = \frac{L^5 f_L^2}{40EI}=\frac{L^5 \Phi_0^2}{40EI}j^2$, where $\Phi_0$ is the flux quantum. This suggests that the coefficient $a_2$ in Eq.(\ref{UaGeneral}) represents the weight of the contribution from the elastic deformation.

The above discussion suggests that if we let $a_n=0$ (n$>$2) in Eq.(\ref{UaGeneral}), then it is equal to ignore the inelastic deformation and interaction of the vortices. The activation energy of the elastic vortices becomes
\begin{equation}
\label{UaQuadratic}
U_a(j) = U_c - a_1 j - a_2 j^2
\end{equation}
substitute Eq.(\ref{UaQuadratic}) into Eq.(\ref{UvsTime}), we have (choose one of the solutions which is a decreasing function of $t$),
\begin{equation}
\label{jElastic}
j(t) = -\frac{1}{2} \frac{a_1}{a_2} \left[1 - \sqrt{1 + 4\frac{a_2}{a_1^2} w(t)}\right] \\
\end{equation}
where $w(t)$ is defined by Eq.(\ref{FunctionW}). Expanding the right hand side of Eq.(\ref{jElastic}) (or using Eq.(\ref{bns}) directly) and keeping the terms up to the second order, we have $j(t) \approx \frac{1}{a_1} w(t) - \frac{a_2}{a_1^3} w^2(t)$. The first order term $\frac{1}{a_1} w(t)$ represents the logarithmic decay. The second order term $\frac{a_2}{a_1^2} w^2(t)$ causes deviation away from the logarithmic decay. Since $\frac{a_2}{a_1^2} w^2(t)$ depends on $a_2$, it represents the contribution from the elasticity of the vortices. This indicates that any elastic vortex under a driving force will show non-logarithmic current decay behavior.

Replacing the $w(t)$ in Eq.(\ref{jElastic}) with $U_c$, we obtain the critical current density $j_c = -\frac{1}{2} \frac{a_1}{a_2} \left[1 - \sqrt{1 + 4\frac{a_2}{a_1^2} U_c}\right]$.

3. \textit{Meaning of $a_1$}.---
Further let $a_2=0$ in Eq.(\ref{UaQuadratic}) is equal to assume that the vortices have very large elastic modulus and elastic deformation can be ignored, i.e., the vortices are rigid. The corresponding activation energy becomes
\begin{equation}
\label{UaLinear}
U_a(j) = U_c - a_1 j
\end{equation}
This is Anderson's linear activation energy \cite{Anderson}. Using Eq.(\ref{JvsW}) and Eq.(\ref{bns}), we obtained the logarithmic current model $j(t) = \frac{1}{a_1} w(t)$. The corresponding critical current density is $j_c = \frac{1}{a_1} U_c$.

Since the Lorentz force on vortices is proportional to the current density $j$, the linear activation energy $U_a(j)$ in Eq.(\ref{UaLinear}) is a result of the Lorentz force causes a direct reduction of the value $a_1 j$ in the pinning potential $U_c$. The early studies have shown that constant $a_1$ is of the order $\Phi_B d^2$, where $\Phi_B$ is the total flux in the flux bundle and $d$ is the length of the bundle.\cite{Anderson,Labusch}

From the above discussion we know that the linear activation energy and corresponding logarithmic decay behavior only occurs in the rigid vortex systems. The elasticity and interaction of vortices modify the linear activation energy, which cause a deviation away from the logarithmic decay behavior. Since the elasticity and interaction of vortices are closely related to the penetration depth $\lambda$ of a superconductor,\cite{Tinkham} we can now analyze the possible effects of $\lambda$ on the flux relaxation phenomenon.

The shear modulus $C_{66}\approx\frac{\Phi_0 B}{(8\pi\lambda)^2}$, uniaxial compression modulus and tilting modulus $C_{11} \approx C_{44} \approx \frac{B^2}{4 \pi}\frac{1}{1+\lambda^2 k^2}$ are all decreasing functions of $\lambda$, where $k$ is the wave vector.\cite{Brandt} Low temperature superconductors have penetration depth $\lambda \sim 500 \dot{A}$, which is smaller than that of high temperature superconductors, $\lambda \sim (1400-2000) \dot{A}$. (Ref.~\onlinecite{Tinkham}) This means that the vortices in low temperature superconductors are stiffer, and there should be a better chance of observing quasi-logarithmic decays in these superconductors.

On the other hand, the penetration depth, $\lambda(T) \approx (1-T/T_c)^{-1/2}$, is an increasing function of temperature $T$. The elastic moduli are then decreasing functions of $T$. At higher temperatures, vortices become soft and undergo large deformations, resulting nonlinear activation energy. Therefore, it is more likely to observe non-logarithmic decay behavior at higher temperatures, but logarithmic decay behavior at lower temperatures.\cite{Ma,Chowdhury}

\section{Conclusion}

The activation energy of flux relaxation phenomenon can be expanded as a Taylor series of current density. The elasticity of the vortices and interaction between the vortices result in the nonlinear activation energy, which causes non-logarithmic decay behavior. It should be emphasized that the derivations in this article were mainly based on mathematical skills. This ensures a wide application field, but the explicit studies on the physical premises should be undertaken according to specific conditions. Secondly, I have neglected the surface pinning effects, which may add significant influence on the flux relaxation phenomenon at higher temperatures close to the critical level. These need to be considered separately.

\section{Acknowledgments}

I thank Aly, David C. Fortin and Yi Shi for carefully reading this manuscript.

\newpage

\textbf{FIGURE CAPTIONS:}
\\
\\
\begin{figure}[htb]
\begin{center}
\includegraphics[width=70mm, height=60mm]{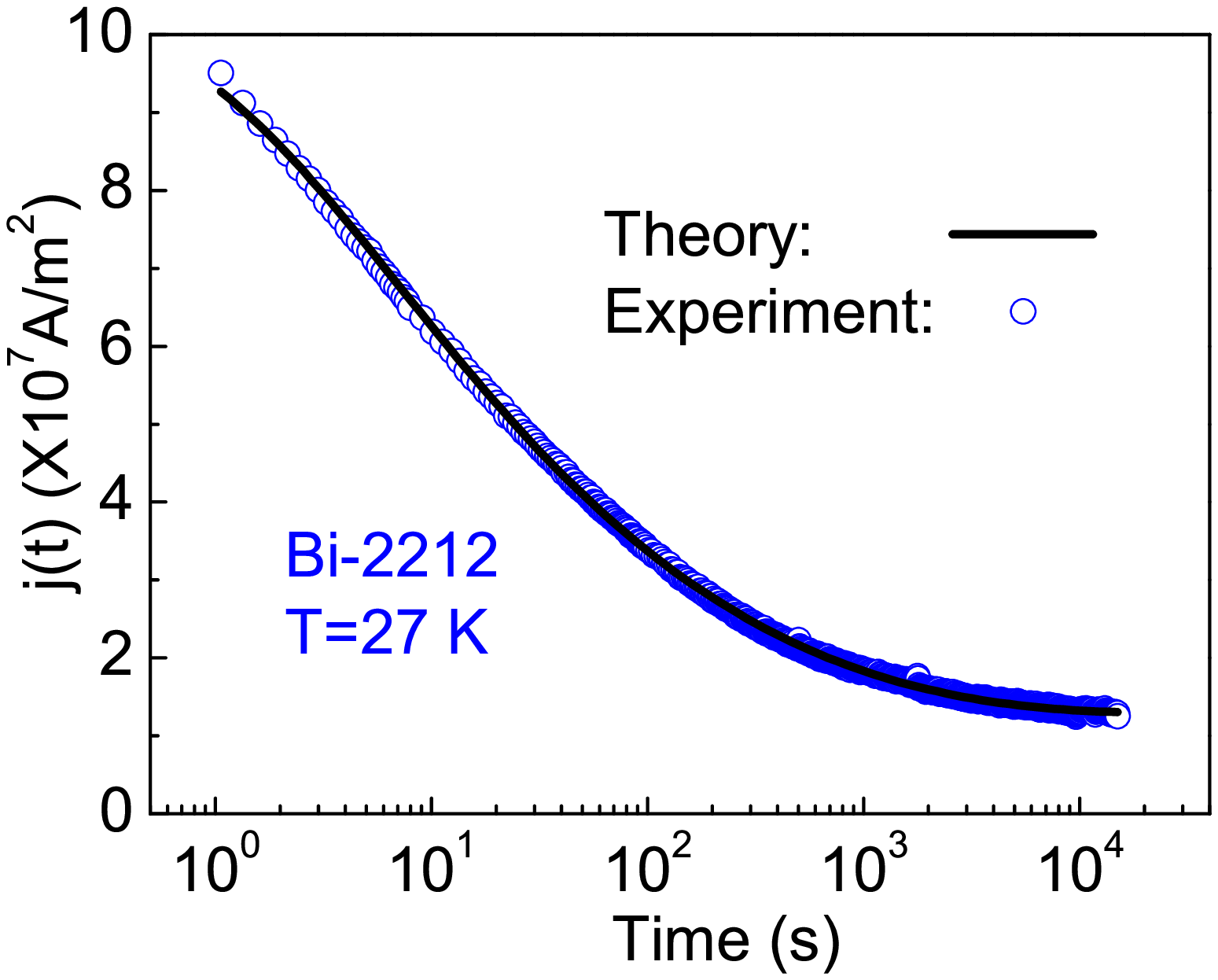}
\caption{\label{figure1} (Color online) Time evolution of persistent current. The scattering points are the experimental data of a $Bi_2Sr_2CaCu_2O_{8+x}$ (Bi-2212) single crystal, i.e., the persistent current induced at 27 K under an applied magnetic field of 750 Gauss.\cite{Ma} The solid black line is the theoretical fit. The fitting results are:
$j(t)=b_1w(t)+b_2w^2(t)+b_3w^3(t)$ (where $w(t)=U_c-27*k*ln(1+t/t_0)$), $b_1=(2.77\pm0.06)\times10^5/k$, $b_2=-(1.79\pm0.01)\times10^3/k^2$, $b_3=(3.63\pm0.08)/k^3$, $U_c=(4.62\pm0.08)\times10^2 k$, $t_0=(1.55\pm0.04)$.}
\end{center}
\end{figure}

\end{document}